\documentclass[prl,superscriptaddress,showpacs,twocolumn]{revtex4}

\usepackage{graphicx}

\usepackage{amssymb,amsfonts,amsmath}
\usepackage[nooneline,raggedright]{subfigure}

\usepackage[dvips]{color}

\newcommand{\fref}[1]{Fig.~\ref{#1}}

\newcommand{\im}{%
           \imath}
\newcommand{\bra}[1]{\ensuremath{\langle #1|}}
\newcommand{\ket}[1]{\ensuremath{|#1\rangle}}

\def\t2g{\ensuremath{t_{2g}}}
\def\a1g{\ensuremath{a_{1g}}}

\newcommand{\svek}{%
        \mathbf}

\newcommand{\vek}[1]{%
        \hbox{\textbf #1}}

\def\etal.{{\it et~al.}}

\def\XXint#1#2#3{{\setbox0=\hbox{$#1{#2#3}{\int}$}
\vcenter{\hbox{$#2#3$}}\kern-.5\wd0}}

\newcommand{\oo}[1]{\frac{1}{#1}}

\begin{document}

\title{Signatures of electronic correlations in iron silicide}

\author{Jan M. Tomczak$^*$}
\affiliation{Department of Physics and Astronomy, Rutgers University, Piscataway, New Jersey 08854, USA}
\author{K. Haule}
\affiliation{Department of Physics and Astronomy, Rutgers University, Piscataway, New Jersey 08854, USA}
\author{G. Kotliar}
\affiliation{Department of Physics and Astronomy, Rutgers University, Piscataway, New Jersey 08854, USA}

\maketitle

{\bf The intermetallic FeSi exhibits an unusual temperature dependence in its electronic and magnetic degrees of freedom,
epitomized by the crossover from a low temperature non-magnetic semiconductor to a high temperature paramagnetic metal 
 with a Curie-Weiss like susceptibility. 
Many proposals for this unconventional behavior have been advanced, 
yet a consensus remains elusive.
Using realistic many-body calculations, we here reproduce the signatures of the metal-insulator crossover in various observables~: 
the spectral function, the optical conductivity, the spin susceptibility, and the Seebeck coefficient.
Validated by quantitative agreement with experiment, we then address the underlying microscopic picture. 
We propose a new scenario in which FeSi is 
a band-insulator at low temperatures and is metalized with increasing temperature 
through correlation induced incoherence. 
We explain that the emergent incoherence is linked to the unlocking of iron fluctuating moments 
which are almost temperature independent at short time scales.
Finally, we make explicit suggestions for improving the thermoelectric performance of FeSi based systems. 
}


Iron based narrow gap semiconductors such as FeSi, FeSb$_2$, or FeGa$_3$ show a pronounced resemblance to heavy fermion Kondo insulators in their
charge 
\cite{Wolfe1965449,PhysRevLett.71.1748,0295-5075-80-1-17008,1742-6596-200-1-012014} and 
spin 
\cite{PhysRev.160.476,1742-6596-200-1-012014} degrees of freedom.
Besides, these systems display a large thermopower
\cite{Wolfe1965449,Buschinger1997784, PhysRevB.83.125209, sun_dalton,0295-5075-80-1-17008,jmt_fesb2}, heralding their potential use 
in solid state devices.
There are two complementary approaches for explaining these unusual properties~: 
On one hand it has been proposed that lattice degrees of freedom play a crucial role
\cite{PhysRevB.59.15002,PhysRevB.83.125209,Delaire22032011}.
On the other hand, electron-electron correlation effects have been invoked on the basis of both
experimental results
\cite{0295-5075-80-1-17008, sun_dalton,PhysRevLett.71.1748,1742-6596-200-1-012014}, as well as
theoretical model studies
\cite{PhysRevB.51.17439,JPSJ.46.1451,PhysRevB.81.125131,PhysRevB.78.033109},
advocating in particular Hubbard physics
\cite{PhysRevB.51.17439,PhysRevB.81.125131,PhysRevB.78.033109},
spin fluctuations 
\cite{JPSJ.46.1451}, spin-state transitions
\cite{PhysRevLett.76.1735,vanderMarel1998138}, or a thermally induced mixed valence
\cite{PhysRevB.50.9952}.

\begin{figure}[!b!]
  \begin{center}
   \includegraphics[angle=-90,width=.49\textwidth]{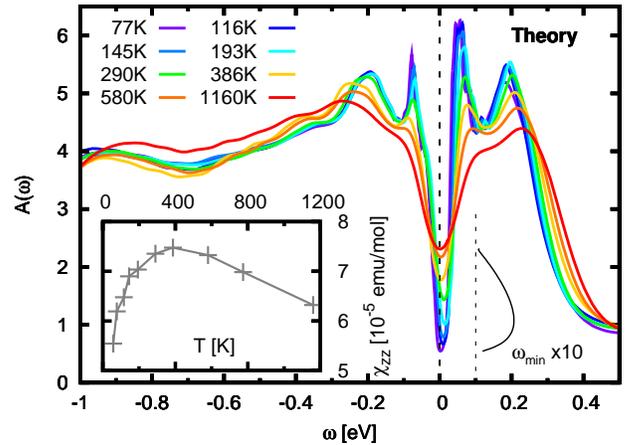}
      \caption{{\bf Local spectra and susceptibility.} The theoretical local spectral function for different temperatures. $\omega_{min}$ traces the spectral minimum with respect to the Fermi level (10x magnified).  Inset~: local spin susceptibility.}
      \label{fig1}
      \end{center}
\end{figure}

Here, we go beyond modelistic approaches and investigate the effect of correlations on prototypical FeSi from the {\it ab initio} perspective.
The key issues that we address are
 (1) can electronic Coulomb correlations alone quantitatively account for the signatures of the temperature induced crossover
in various observables, and
 (2) what is the underlying microscopic origin of this behavior~?
As a realistic many-body approach, we employ the combination of density functional theory and dynamical mean-field theory DFT+DMFT (for a review see e.g.\ \cite{RevModPhys.78.865})
as implemented in Ref.\cite{PhysRevB.81.195107}.
For the calculation of the Seebeck coefficient we have extended our previous work\cite{jmt_fesb2} for DFT computations to include the DMFT self-energy,
in a full orbital setup. For details see the supporting information.

At low temperatures, iron silicide
is a semiconductor with a gap $\Delta\approx50-60$meV
\cite{PhysRev.160.476,PhysRevLett.71.1748,0295-5075-28-5-008}, with
the resistivity 
\cite{Wolfe1965449,Buschinger1997784} 
 and the magnetic susceptibility 
 \cite{PhysRev.160.476} 
following activation laws.
At 150-200K (i.e.\ at temperatures much smaller than $\Delta$)
a crossover to a (bad) metal is observed in transport 
\cite{Wolfe1965449,Buschinger1997784,PhysRevB.56.12916}, and optical spectroscopy 
\cite{PhysRevLett.71.1748,0295-5075-28-5-008,PhysRevB.55.R4863,vanderMarel1998138,PhysRevB.56.12916,PhysRevB.79.165111}.
 Moreover, FeSi displays a maximum in the susceptibility at 400K, followed by a Curie-Weiss like law 
\cite{PhysRev.160.476}.
The energy scale over which spectral weight in the optical conductivity is transfered through the transition
has been long debated
\cite{PhysRevLett.71.1748,vanderMarel1998138,0295-5075-28-5-008,PhysRevB.56.12916}.  
Recent ellipsometry measurements 
\cite{PhysRevB.79.165111} showed that 
weight is moved over several eV  -- a common harbinger of correlation effects
\cite{PhysRevB.54.8452}.
The Seebeck coefficient of FeSi peaks near 50K with a remarkable 700$\mu$V/K,  
but is quickly suppressed at higher temperatures, when the unusual spin and charge properties set in 
 \cite{Wolfe1965449,Buschinger1997784,PhysRevB.83.125209}.

\section{Spectral Properties}
\fref{fig1} shows our theoretical local spectral function of FeSi at various temperatures
[orbital and momentum resolved spectra can be found in the supporting information].
At low temperature, the spectrum is similar to that obtained within band-theory
\cite{PhysRevB.47.13114,PhysRevB.59.15002,PhysRevB.81.125131},
albeit with a gap renormalized by a factor of about 2,  in agreement with photoemission spectroscopy (PES)
\cite{PhysRevLett.101.046406,PhysRevB.77.205117}.
\footnote{From the slope of the self-energy $\Sigma$, we extract m*/m=1.5, the rest of the gap shrinking comes from inter-orbital shifts as induced by the real parts of the self-energy $\Re\Sigma(0)$.}
The electronic excitation spectrum is thus band-like and coherent at low temperature.
As was noted earlier
\cite{PhysRevB.47.13114,PhysRevB.59.15002,PhysRevB.81.125131}, the gap edges are very sharp in this regime,
 indicating a potentially large thermopower
\cite{Mahan07231996}.
At higher temperatures, features broaden, and the system becomes a bad metal as found experimentally
\cite{PhysRevLett.101.046406,PhysRevB.77.205117}.
We stress that this effect is largely beyond a mere temperature broadening of the electron distribution (Fermi) function
\cite{PhysRevLett.71.1748,PhysRevLett.101.046406,PhysRevB.77.205117}.
Due to the asymmetry of the spectrum, the position of the gap moves with temperature~:
In agreement with PES
\cite{PhysRevLett.101.046406,PhysRevB.77.205117} the minimum $\omega_{min}$ of the spectrum [depicted in \fref{fig1}]
starts out near the middle of the gap (as expected for a semiconductor\cite{jmt_fesb2}), then
first moves up with increasing temperature.
Above 300K, $\omega_{min}$ again approaches the Fermi level, as the asymmetry is reversed.

\section{Optical Spectroscopy}
\begin{figure}[!h!]
  \begin{center}
   \includegraphics[angle=-90,width=.49\textwidth]{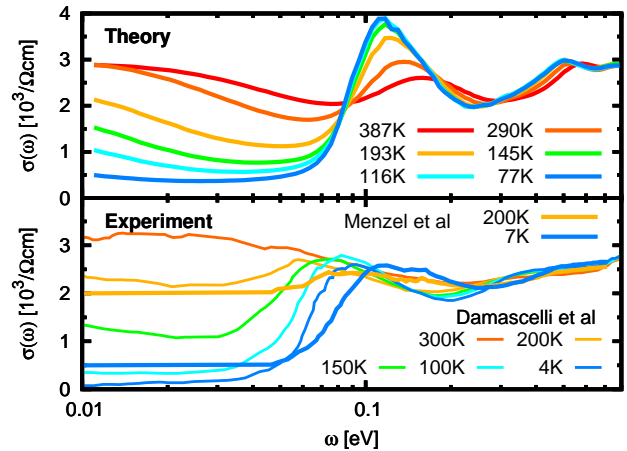}
      \caption{{\bf Optical conductivity.} Theoretical (top) and experimental \cite{PhysRevB.79.165111,PhysRevB.55.R4863} (bottom) optical conductivity as a function of frequency for various temperatures.}
      \label{fig2}
      \end{center}
\end{figure}

\begin{figure}[!h!]
  \begin{center}
   \includegraphics[angle=-90,width=.49\textwidth]{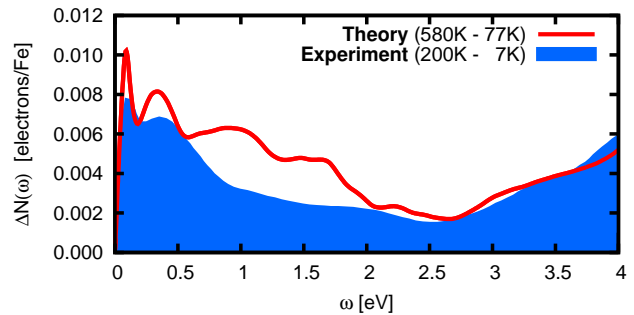}
     \caption{{\bf Transfer of spectral weight.} Difference of high and low temperature carrier density as obtained from the integrated spectral weight [see text for details]. Experimental results from \cite{PhysRevB.79.165111},  used temperatures as indicated.}
      \label{fig3}
      \end{center}
\end{figure}

Owing to a high precision, and the existence of sum-rule arguments, optical spectroscopy is a valuable tool for tracking the evolution
of a system under change of external parameters. 
Transfers of spectral weight over scales related to the Coulomb interaction $U$ rather than the gap $\Delta$ are usually considered a hallmark
of correlation effects
\cite{PhysRevB.54.8452}.  
Recent ellipsometry measurements
\cite{PhysRevB.79.165111} showed spectral weight transfers over several eV.
In \fref{fig2} we compare our realistic optical conductivities $\sigma(\omega,T)$ to experiment 
\cite{PhysRevB.79.165111,PhysRevB.55.R4863}. 
Both, the overall magnitude and the crucial temperature evolution are well captured~: Cooling down from $400$K depletes
spectral weight below 80meV, with only parts of it being transfered to energies just slightly above the gap.
To analyze this in more detail, we follow Menzel et al.\cite{PhysRevB.79.165111} and plot in \fref{fig3} the temperature difference $\Delta N(\omega)=N_{T_1}(\omega)-N_{T_2}(\omega)$ of the effective number of carriers
 $N_T(\omega)=\frac{2m_e V}{\pi e^2}\int_0^\omega d\omega^\prime \sigma(\omega^\prime,T)$ as a function of energy.
An intersection with the x-axis corresponds to a full recovery of spectral weight as is imposed by the f-sum rule.
Our theoretical results quantitatively trace the experimental temperature dependence.
We note that there are several isosbectic points in the optical conductivity, \fref{fig2}, which lead to extrema in $\Delta N(\omega)$ in \fref{fig3}.
The first peak in $\Delta N$ is at 80meV, the scale of the semiconducting gap, above which spectral weight starts to pile up at low temperature in $\sigma(\omega)$.
The first minimum in $\Delta N(\omega)$ occurs at the second isosbectic point at around 0.18eV, up to where only $\sim 35\%$ compensation of excess carriers is achieved for the
theoretical curve.
As is clear from \fref{fig3}, $\Delta N(\omega)$ does not vanish over the extended energy range plotted, hence a total compensation is not reached below the scale of the Coulomb repulsion of 5 eV. 
As a further assessment, we show, in \fref{respf}(a), the theoretical resistivity in comparison with several experiments.

\section{Thermopower}
FeSi boasts a notably large thermopower at low temperatures
\cite{Wolfe1965449,Buschinger1997784, PhysRevB.83.125209}. 
Yet, it is compatible in magnitude with a Seebeck coefficient of purely electronic origin~: For a band-like semiconductor in the regime $k_BT<\Delta$, the electron diffusive thermopower
cannot exceed $\Delta/T$\cite{jmt_fesb2}. From this perspective, a larger gap favors the thermopower.
While validated in FeSi, the above constraint is violated in the related compound FeSb$_2$\cite{0295-5075-80-1-17008, sun_dalton,jmt_fesb2}.
Given the dominantly electronic picture, in conjunction with the band-like nature of FeSi at low temperature, 
the Seebeck coefficient in that regime can be accounted for by band theory~:
Indeed, it was shown that a slightly hole doped band-structure yields good agreement below 100K
\cite{PhysRevB.59.15002}.
As shown in \fref{fig4}, we confirm this by supplementing band-theory with an effective mass of 2 and 0.001 holes/Fe.
While a dependence on stoichiometry is seen in experiments\cite{Buschinger1997784}, the tiny amounts of extra holes should be viewed as a means to alter the particle-hole asymmetry
\cite{jmt_fesb2} rather than an effect of excess charge.
As is well known,
the particle-hole asymmetry plays a major role in determining the thermopower, since electron and hole contributions have opposite signs.
The convention is that a negative Seebeck coefficient is dominated by electron transport, and a positive one by holes.
It is interesting to note that for an insulator a large thermopower is expected {\it near} thermoelectric particle/hole symmetry\cite{jmt_fesb2}, with a
large sensitivity to the exact imbalance of carriers.
Scanning through electron and hole doping, FeSi is experimentally indeed placed near such a  
boundary
\cite{JPSJ.76.093601}.

As expected, introducing an effective mass alone fails at higher temperatures. 
It has been conjectured (but not calculated) that 
this could be
accounted for by thermal disorder via the electron-phonon coupling
\cite{PhysRevB.59.15002, Delaire22032011}.
On the other hand, model studies indicated compatibility with the picture of Coulomb correlations
\cite{JPSJS.71S.288}.
In \fref{fig4} we display the Seebeck coefficient obtained from our realistic many-body calculation. The agreement  with experiments is very good. 
Notably, the theory captures the overall suppression of the thermopower above 100K,
which comes from an enhanced conductivity due to the accumulation of incoherent weight at the Fermi level.
In this sense incoherence is detrimental to semi-conductor based thermopower. Note however that correlation effects can substantially enhance the Seebeck coefficient of 
correlated metals [see e.g.\ \cite{PhysRevB.65.075102}].

The temperature dependence of the Seebeck coefficient in \fref{fig4} is connected to the moving of the chemical potential discussed above~:
As a function of rising temperature, the chemical potential first moves down, thereby increasing the electron contribution to the thermopower
\cite{jmt_fesb2}.
At some temperature thermoelectric particle-hole symmetry is passed, and the Seebeck coefficient is dominated by electrons ($S<0$), while at higher
temperatures, the trend is reversed, and $S$ becomes positive again.

\begin{figure*}[!t!]
\centering
\subfigure[]{\scalebox{1.}{\includegraphics[angle=-90,width=0.435\textwidth]{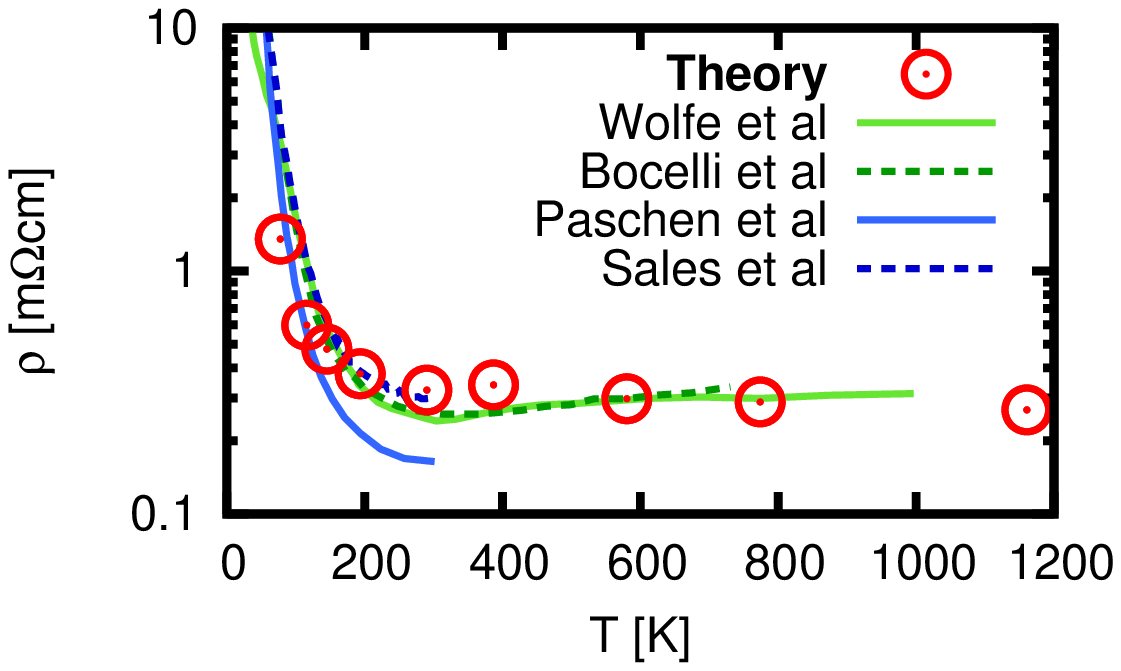}}}
\hspace{1.25cm}
\subfigure[]{\scalebox{1.}{\includegraphics[angle=-90,width=0.435\textwidth]{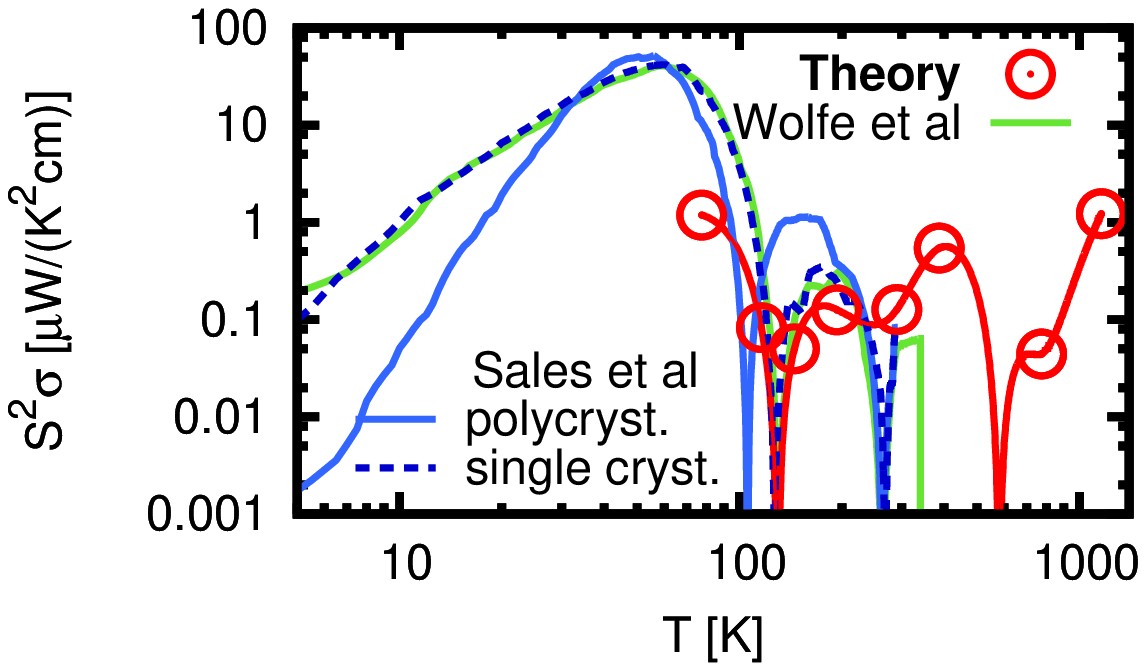}}}
\caption{{\bf Resistivity and Powerfactor.} (a) Shown is a comparison of the theoretical resistivity of FeSi with several experimental results\cite{Wolfe1965449,Bocelli_FeSi,PhysRevB.56.12916,PhysRevB.83.125209}. (b) the powerfactor of FeSi, theory in comparison
with experimental data compiled from \cite{Wolfe1965449,PhysRevB.83.125209}.}
\label{respf}
\end{figure*}

\begin{figure}[!t!]
  \begin{center}
   \includegraphics[angle=-90,width=.49\textwidth]{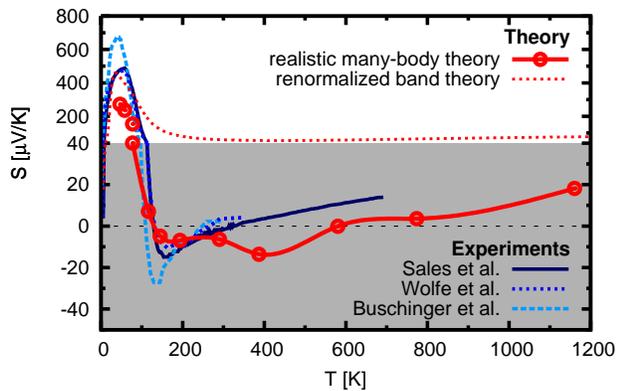}
      \caption{{\bf Thermopower.} Theoretical Seebeck coefficient in comparison with experiment
      \cite{Wolfe1965449,Buschinger1997784,PhysRevB.83.125209}. Note the change in scale above 40$\mu$V/K. The line connecting the many-body theory results is a guide to the eye. The unconnected points at low temperature are obtained from an analysis of the asymptotic behavior of the many-body calculation.}
      \label{fig4}
      \end{center}
\end{figure}

We note that the conversion efficiency of thermoelectric devices is measured by the so-called figure of merit 
$ZT=S^2\sigma T/\kappa$. Here, the combination $S^2\sigma$ of the Seebeck coefficient $S$ and the conductivity $\sigma$ is called the power factor (PF),
and $\kappa$ is the thermal conductivity.
The power factor of FeSi, displayed in \fref{respf}(b),  peaks at around 60K, where is reaches more than 40$\mu W/(K^2cm)$, i.e.\ it reaches values
comparable to state of the art Bi$_2$Te$_3$ (at its maximum value at $550$K), and is for $T\ge 60$K larger than in
FeSb$_2$ that holds the overall record PF (realized at 12K)\cite{0295-5075-80-1-17008}.

\section{Microscopic Insights}
The quantitative agreement of our theoretical results with a large panoply of experimental data validates our approach, thus signaling the paramount influence of electronic correlation effects.
Drawing from the microscopic insights of our method, we now address the physical picture underlying the intriguing properties of FeSi.
This will in particular allow us to propose ways to improve the thermoelectrical properties of FeSi.

\subsection{Crossover to the Metallic State}
Within our theoretical picture, the crossover to bad metallic behavior is {\it not}
caused by a narrowing of the excitation gap [see the supporting information for details].
Instead, it is filled with incoherent weight that emerges with increasing temperature.
Information about the coherence of the one-particle excitations are encoded in the imaginary parts of the electron 
self-energy $\Sigma$. For the relevant orbital components 
 we find
 $\Im\Sigma(\omega=0)\approx -aT^2$ with $a=1.9\cdot 10^{-4}$meV/K$^2$. $\Im\Sigma$ thus reaches the value of $\Delta/2$ at around 400K,
 when only a pseudogap remains [see \fref{fig1}].

Complementarily, it was proposed
that the arguably large electron-phonon
coupling 
\cite{PhysRevB.79.165111, PhysRevB.83.125209,Delaire22032011} causes the closure of the gap
via thermal induced atomic disorder 
\cite{PhysRevB.59.15002,Delaire22032011}. In molecular dynamics simulations
\cite{Delaire22032011},
the gap $\Delta_{DFT}$ was shown to vanish abruptly for temperatures of the order of $T\approx\Delta_{DFT}/2$,
in contrast to the gradual transition that is observed in experiment and reproduced by our theory.
We note that also in other systems with large electron-phonon coupling, spectral weight transfers are quantitatively accounted for by
electronic correlations\cite{optic_prb}.

\subsection{Strength of Correlations}
A recurring question in condensed matter physics is whether a system is well described in either an itinerant or a localized picture. 
Both the low effective mass of FeSi ($m^*/m$$\approx$$1.5$) and the rather high
kinetic energy $E_{kin}$$\approx$$-10.5$eV of the iron states signal a large degree of delocalization.
Indeed, other iron compounds show significantly higher masses and lower kinetic energies, 
e.g.\ the pnictides BaFe$_2$As$_2$  ($E_{kin}$[eV];$(m^*/m)_{xy})$$\approx$$(-7;3)$ and CaFe$_2$As$_2$ $(-8; 2.5)$ or the chalcogen FeTe $(-6.5; 7)$\cite{Yin_pnictide}.
Accordingly, FeSi is an only moderately correlated, itinerant material.
Yet, effects of correlation induced incoherence are essential for the crossover to the metallic state.
This seeming contradiction is resolved by noting
that all relevant energy scales are of similar magnitude.
 Indeed~: $\mathcal{O}(\Delta)\approx\mathcal{O}(\Im\Sigma)\approx \mathcal{O}(T)\approx\mathcal{O}(50\hbox{meV})$, 
 which leads low energy properties to be correlation dominated.

\subsection{The Spin State}
A major signature of the unconventional behavior of FeSi is the non-monotonous uniform magnetic susceptibility
\cite{PhysRev.160.476}. 
The latter is closely mimicked by our local spin susceptibility $\chi_{loc}(\omega=0)\sim\frac{1}{\beta}\int d\tau \langle S_z(\tau)S_z(0)\rangle$ [see inset of \fref{fig1}],
indicating compatibility with the picture of electronic correlations. 
Interestingly there is a distinction of time scales~:
While $\chi_{loc}(\omega=0)$ -- the time-averaged response -- displays a strong temperature dependence, the spin response at very short time scales, as probed
by the observable $\lim_{\tau\rightarrow 0}\langle S(\tau)S(0)\rangle$, is virtually temperature independent. From the latter, we obtain
an effective moment $M=\sqrt{S(S+1)}g_s\approx3$ ($g_s=2$).
This is consistent with major contributions from effective iron states with $S=1$ [see supporting information for details], and in agreement with 
$M=2.7$ as obtained from fitting the experimental susceptibility 
\cite{PhysRev.160.476} 
to a Curie-Weiss law $\chi=\frac{\mu_0\mu_B}{3k_B}M^2/(T-T_C)$ for $T>400$K.
%
We note that the local susceptibility [see inset of \fref{fig1}] is about one order of magnitude smaller than the experimental uniform susceptibility.
Hence, the temperature induced fluctuating moment of the underlying spin state is, to a large extent, not local.
This is corroborated by neutron experiments\cite{PhysRevLett.59.351} that find a significant magnetic scattering at ``ferromagnetic'' reciprocal lattice vectors.

Since the effective moment $M$ is constant, FeSi thus does not undergo a spin-state transition
as is found e.g.\ in MnO  
or LaCoO$_3$. 
There it originates from
 a competition between the Hund's coupling $J$ and the crystal field splittings.
In FeSi, the splitting of low energy excitations is given by the gap $\Delta$, which is much smaller
than the Hund's coupling $J$,
thereby favoring the high spin configuration.

The preponderance of S=1 states in particular implies that FeSi is not a singlet insulator as previously proposed
\cite{PhysRevLett.76.1735,vanderMarel1998138}.
We further find that FeSi is in a mixed valence state. We obtain an iron valence of $N_d\approx6.2$, with 
a large variance $\delta N=\langle (N-\langle N\rangle)^2\rangle\approx 0.93$ [see supporting information for details].
However, these numbers are  insensitive to temperature,
thus excluding a thermally induced mixed valence
\cite{PhysRevB.50.9952}.
%

Interestingly, we find that our results are one order of magnitude more sensitive to the strength of the Hund's rule coupling $J$ than to the Hubbard interaction $U$ [for details see the supporting information].
This effect is reminiscent of the physics of Hund's metals as found in iron pnictides and chalcogenides\cite{1367-2630-11-2-025021,Yin_pnictide}, and is quite different from Hubbard physics in which the Hubbard $U$
reduces charge fluctuations and the time-scale associated with them, while the time averaged local spin susceptibility is strongly enhanced.
In contrast, only the spin fluctuations but not the charge fluctuations are short-lived in FeSi, resulting in only a moderate enhancement of the (time averaged) local spin susceptibility, but a strongly enhanced short time (energy averaged) fluctuating magnetic moment.

\section{The Physical Picture}
We are now in the position to elucidate the fundamental picture of FeSi~:
(a) What is the physical origin of the crossover in the susceptibility~? (b) How is the latter linked to the emergence of incoherent spectral weight, i.e.\ 
what is the relation between the crossovers for the spin and charge degrees of freedom~?

At low temperatures, FeSi is a conventional band-like semiconductor, i.e.\ it is approximately described by an effective one-particle Hamiltonian that can
be diagonalized in momentum space. Excitations are well-defined (coherent). The system is in a high spin state, but spin excitations are gapped,
and so the spin susceptibility is small as fluctuations at finite time scales are quenched. In this sense the spin degrees of freedom are inactive.
With increasing temperature, however, a fluctuating moment develops at the iron sites.
The emergence of such a locking to the real-space lattice breaks down the momentum-space description~: $\mathbf{k}$ is no longer a good quantum number, hence excitations
acquire a finite lifetime broadening, and the spectrum becomes incoherent.

Out theory goes beyond previous theoretical proposals, indeed it reconciles (and validates by including explicitly the Si degrees of freedom) two seminal model-based approaches~:
The model of Fu and Doniach\cite{PhysRevB.51.17439} that addressed the evolution of the one particle spectrum, and which can serve as a simplified cartoon of our realistic
calculation, and the  
spin-fluctuation theory of Takahashi and Moriya\cite{JPSJ.46.1451} which is compatible with our results for the spin degrees of freedom.
Here, we claim that the crossovers in the spin and charge response are actually intimately linked to each other.

Due to the complicated multi-orbital nature of the problem, there are however important differences from previous model calculations\cite{PhysRevB.51.17439,PhysRevB.81.125131,PhysRevB.78.033109}~:
In the realistic case the relevant control parameter is the Hund's coupling $J$ rather than the Coulomb interaction $U$ [see the supporting information].
Furthermore, we account for the particle/hole asymmetry.
Indeed, for a symmetric model \cite{PhysRevB.51.17439}, the Seebeck coefficient -- a key feature of this system -- is zero at all temperatures,
and the chemical potential is pinned to its value at 0K.

\section{Conclusions \& Outlook}
To summarize, we have obtained a fundamental microscopic
theory of iron silicide, the canonical example of
a correlated insulator. This material features a competitive power factor in the temperature range of 50-100K.
Our theory suggests concrete ways to improve that already remarkable thermoelectric performance.   First, as is becoming standard practice in thermoelectric development
\cite{Snyder2008}, nano-- or heterostructuring 
is needed to reduce thermal conductivity
to convert the good power factor into a large figure of merit. More important yet, the power factor can be improved further by increasing the charge gap, through a reduction of the ratio of Hund's coupling over bandwidth, or an increase in the inter-atomic hybridizations that cause the band-gap.
To achieve this goal, we propose experimental studies of FeSi under external pressure\cite{Bauer1997794}, or by compressing the lattice using a suitable substrate. A particularly interesting option would be partial iso-electronic substitution of iron with ruthenium\cite{PhysRevB.65.245206}.
Not only does the iso-structural RuSi have a larger gap than FeSi, and itself a notable Seebeck coefficient above 100K\cite{Hohl199839}, but also the fluctuating moment physics that drives the metalization and the quenching of the thermopower in FeSi could be effectively reduced, as well as the thermal conductivity decreased via the alloying.
To our knowledge these avenues have not been pursued vis-{\`a}-vis their impact onto thermoelectricity.
More generally, the theory outlined here, describes the subtle interplay of electron--electron correlations and thermoelectricity, adding a new host of systems amenable to theory assisted thermoelectric material design.

\section{Acknowledgments}
We thank P. Sun and F. Steglich for interesting discussions as well as for sharing unpublished results.
The authors were supported by the NSF-materials
world network under grant number NSF DMR 0806937 and NSF DMR 0906943, and by the PUF program.
Acknowledgment is also made to the donors of the American Chemical Society Petroleum Research Fund 48802 for  partial support of this research.
 
\section{author contributions.}
JT lead the project. JT developed the algorithms and codes to carry out
the ab initio computation of the thermoelectric power within the LDA+DMFT code written by KH. JT carried out the all computations, the initial  interpretation of the results and the writing of the manuscript. All the authors JT, KH and GK,
were involved  in the discussion of the results and participated in the writing of the final draft. 





\clearpage
\newpage




\noindent
{\bf Signatures of electronic correlations in iron silicide -- Supporting Information}\\
Jan M. Tomczak\\
K. Haule\\
G. Kotliar


\section{Method~: Electronic structure}

For the density functional theory (DFT) part, we employ Wien2k\cite{wien2k} within the generalized gradient approximation(GGA)\cite{PhysRevLett.77.3865}.
The dynamical mean field theory (DMFT)\cite{RevModPhys.78.865} impurity problem is solved using a
hybridization expansion continuous time quantum monte carlo (ctqmc) method \cite{PhysRevB.75.155113,PhysRevLett.97.076405}. 
We use a projection based DFT+DMFT setup with full charge self-consistency, as implemented in Ref.\cite{PhysRevB.81.195107}.
For the Coulomb interaction and Hund's coupling, we use the value $U=5.0$eV, $J=0.7$eV, respectively, which have been found appropriate
for iron based compounds\cite{PhysRevB.82.045105}. For the dependence of results on $J$, see the discussion in the extended results section, below.

\section{Method~: Thermopower calculations}

The electronic contribution to the thermopower along the $\alpha$-direction, i.e. the diagonal element $S_\alpha=S_{\alpha\alpha}$ of the Seebeck tensor, can be written as 
(see e.g. Ref.\cite{PhysRevB.67.115131,oudovenko:035120,haule_thermo,jmt_fesb2,Held_thermo})
\begin{eqnarray}
S_{\alpha}&=&-\frac{k_B}{\left|e\right|}  \frac{A_1^{\alpha}}{A_0^{\alpha}}
\label{eqS}
\end{eqnarray}
where $A_{0/1}^\alpha$ are the coefficients that relate (within linear response theory) the perturbed charge/heat current
to the external perturbing field. They can be expressed as
\begin{eqnarray}
A_n^\alpha&=&\int d\omega \beta^n(\omega-\mu)^n \left(-\frac{\partial f_\mu}{\partial \omega}  \right) \Xi_\alpha(\omega)
\label{A01}
\end{eqnarray}
with the Fermi level $\mu$ and the Fermi function $f_\mu$. When neglecting vertex corrections 
in the DMFT spirit\cite{PhysRevLett.64.1990,bible},
the transport kernel can be written as
\begin{eqnarray}
\Xi_\alpha(\omega)&=&\sum_{\vek{k}} \textrm{Tr}\left[ v_\alpha(\vek{k})A(\vek{k},\omega)v_\alpha(\vek{k})A(\vek{k},\omega)\right]
\end{eqnarray}
where the transition matrix elements (Fermi velocities) $v_\alpha(\vek{k})$ and the spectral functions 
$A(\vek{k},\omega)$ are matrices in orbital space.
We choose to evaluate the trace in $\Xi$ in the Kohn-Sham basis $\left\{\Psi_{\svek{k}n}(\vek{r})\right\}$.
Accordingly, the Fermi velocities -- that in the dipole approximation are matrix elements of the momentum operator $\mathcal{P}$ -- are computed as
\begin{eqnarray}
v_\alpha^{nn^\prime}(\vek{k})&=&\frac{1}{m} \bra{\vek{k}n}\mathcal{P}_\alpha\ket{\vek{k}n^\prime} \nonumber\\
&=& -\im\frac{\hbar}{m}\int d^3r \Psi^*_{\svek{k}n}(\vek{r})\nabla_\alpha\Psi_{\svek{k}n^\prime}(\vek{r})
\end{eqnarray}
In wien2k\cite{wien2k} the Kohn-Sham orbitals are expressed in augmented plane waves (and local orbitals). Thus the above integral
has contributions from within muffin tin spheres, interstitials and mixed terms. For details see Ref.\cite{AmbroschDraxl20061}. 
The spectral functions are given by
\begin{eqnarray}
A(\vek{k},\omega)&=&-\oo{2\pi\im} \left[G(\vek{k},\omega)-G^\dag(\vek{k},\omega) \right]
\end{eqnarray}
with the Greens function $G$ in the Kohn-Sham basis
\begin{eqnarray}
\left(G^{-1}\right)_{mm^\prime}(\vek{k},\omega)&=&(\omega+\mu-\epsilon_{\svek{k}m})\delta_{mm^\prime}-  \bar{\Sigma}_{mm^\prime}(\vek{k},\omega) \nonumber \\
\end{eqnarray}
where $\epsilon_{\svek{k}m}$ are the Kohn-Sham energies. $\bar{\Sigma}=\Sigma-E_{DC}$ is the double counting corrected self-energy obtained by embedding the local impurity self-energy $\Sigma^\tau(\omega)$ of atom $\tau$
into the solid by (see Ref. \cite{PhysRevB.81.195107} for further details)
\begin{eqnarray}
{\Sigma}_{nn^\prime}(\vek{k},\omega)&=&\sum_{\tau LL^\prime} P_{\svek{k}}^\tau(nn^\prime,LL^\prime) \Sigma^\tau_{LL^\prime}(\omega)
\end{eqnarray}
where $L,L^\prime$ index the impurity space, in our case referring to spherical harmonics $L=(l,m)$. Here, $P$ is the projection operator that defines the Hilbert space of the correlated orbitals,
as well as the embedding procedure
\cite{PhysRevB.81.195107}. In our DFT+DMFT calculation, we employ full charge self-consistency, whereby not only the spectral functions are influenced by the correlation effects (via the self-energy), but also the transition matrix elements are altered by means of a self-consistent charge density.

\section{Further details of the Results}

\medskip

\paragraph{Band-structure.}
FeSi crystallizes in the B20 structure with the cubic space group P2$_1$3, and four iron atoms per unit cell.
The ligand arrangement splits the 3d orbitals into 1/2/2 for which we find the corresponding major characters  $z^2$/$x^2$-$y^2+ xy$/$xz + yz$, with the latter two multiplets separated by an avoided crossing. With a nominal iron valence of 6, FeSi is an insulator within band theory\cite{PhysRevB.47.13114,PhysRevB.59.15002,PhysRevB.81.125131}. 
Owing to the hybridization nature of the gap, and the absence of bandwidth narrowing effects --and contrary to most other semi-conductors--, DFT calculations overestimate the gap by a factor of about 2~:
$\Delta_{DFT}\approx0.11$eV\cite{PhysRevB.47.13114,PhysRevB.59.15002,PhysRevB.81.125131}.

\medskip

\paragraph{Orbital and momentum resolved Spectra and Influence of Incoherence.}

\begin{figure}[th]
  \begin{center}
{\includegraphics[clip,angle=0,width=.47\textwidth]{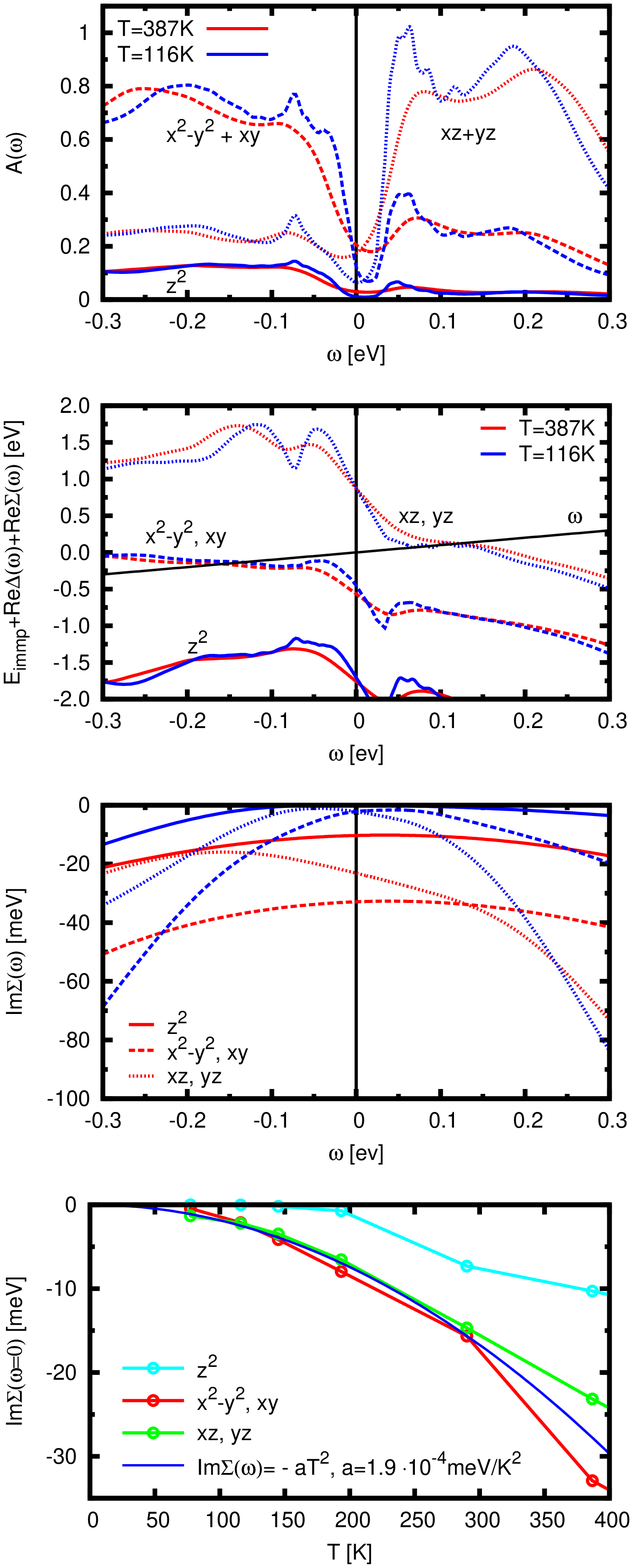}}
      \caption{(a)-(c) Spectral function, real, and imaginary parts of the DMFT self-energy resolved into orbital components are shown for the two temperatures T=116K (blue) and T=387K (red),
      (d) imaginary parts of the self-energy at zero frequency as a function of temperature.
      }
      \label{figSUP_spec}
      \end{center}
\end{figure}

In \fref{figSUP_spec} (a) we display the DFT+DMFT spectral function, resolved into orbital characters of the Fe 3d manifold (different line types)
and for two temperatures (red and blue). The gap (or pseudo gap, depending on the temperature) is formed mainly by orbitals 
of $x^2-y^2$ and $xy$ character on the valence, and $xz$ and $yz$ on the conduction side.
Besides the moving of the chemical potential [see discussion in the main text], the spectral weight at the Fermi level
is increased at larger temperatures by effects of incoherence rather than a closing of the gap though a shifting of excitations. 
In the following, we shall justify this claim.

The impurity Greens function (of atom $\tau$ and orbital $L=(l,m)$) is given by 
\begin{eqnarray}
G^\tau_{LL}(\omega)&=&[\omega-(E^{imp}_L+\Delta^\tau_{LL}(\omega)+\Sigma^\tau_{LL}(\omega))]^{-1}
\end{eqnarray}
where $E_{imp}$ are the impurity levels, $\Delta(\omega)$ the hybridization function, and $\Sigma(\omega)$ the self-energy, and we have assumed all quantities to be diagonal
in the local basis (for details see e.g.\ \cite{PhysRevB.81.195107}). 
A good idea about the major features in the excitation spectrum can be gained (see e.g.\ Ref.\cite{tomczak_vo2_proc}) by analyzing the poles of the Greens function in the absence of lifetime effects, i.e.\ by 
\begin{eqnarray}
\omega&=&E^{imp}_L+\Re\Delta^\tau_{LL}(\omega)+\Re\Sigma^\tau_{LL}(\omega)
\label{eqQP}
\end{eqnarray}
which is shown in \fref{figSUP_spec} (b) where the intersections with the black line mark the graphical solutions for the above equation.
As expected, poles in the occupied part are of $x^2-y^2$ and $xy$ character, whereas as for $\omega>0$ solution can be found for the $xz$ and $yz$ components.

The interesting observation here is that those poles slightly move further {\it apart} when the temperature is {\it raised}, i.e.\ the gap in the quasi-particle band-structure
actually becomes larger. Yet at the same time, as is evident from \fref{figSUP_spec} (c), the imaginary part of the self-energy increases drastically.
The values of $\Im\Sigma$ are seemingly small, but they are comparable in magnitude to the experimental charge gap ($50-55$meV).
\fref{figSUP_spec}(d) shows that the values at the Fermi level of the relevant orbital components of $\Im\Sigma$ follow up to 400K a $T^2$ law with a coefficient $a=-1.9 \cdot 10^{-4}$meV/K$^2$.
Assuming Lorentzian line shapes, the combined
half-width-at-half-maximum  of the conduction and valence bands 
reaches the size of the gap below 400K~: $2\Im\Sigma(\omega=0,T=363\hbox{K})=50$meV.
Yet, we note that the filling of the gap is gradual and already at lower temperatures, there appears finite incoherent weight at the Fermi level.
This is different from the closure of the gap in the picture of thermal disorder\cite{Delaire22032011}, mentioned in the manuscript, 
where the bridging of the gap occurs rather abruptly for $T\approx\Delta/2$.
Also, due to the particle/hole asymmetry, the chemical potential falls into the middle of the gap only at T=0 [see also the main text, and Fig.1 there].
This further lowers the temperature at which sizable incoherent weight appears at low energy.
%

For accessing lower temperatures in the Seebeck coefficient than is possible for the Monte Carlo to reach (see the main manuscript, and, there, the unconnected points in Fig.~4), we
extrapolate the imaginary parts of the self-energy of the run at the lowest temperature following the $T^2$ behavior described above.

\begin{figure*}[h!b]
  \begin{center}
\includegraphics[clip,angle=0,width=0.95\textwidth]{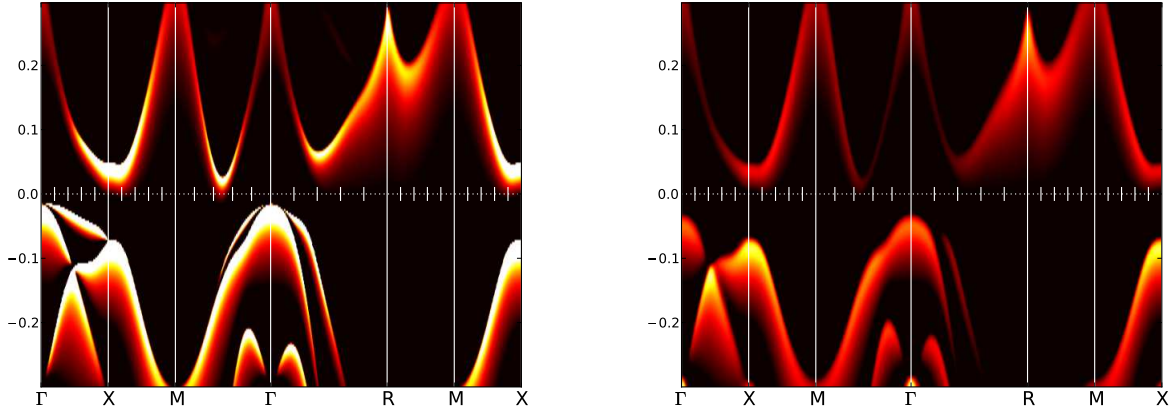}
      \caption{Momentum resolved many-body spectral function at $T=116$K (left) and $T=387$K (right). Energies in eV, the Fermi level is at the origin.}
      \label{figSUP_akw}
      \end{center}
\end{figure*}

In \fref{figSUP_akw}, we further show the total spectral function momentum-resolved along various symmetry lines, and for two temperatures.
As was evident already from the local spectra, excitations broaden significantly with temperature. 
Also seen is the slight moving apart of the peak intensities at higher T, i.e. the widening of the gap 
as given by the poles of the Greens function discussed above. While the incoherent spectral weight at the Fermi level is not discernible on the used color scale,
it is evident that the crossover to a (bad) metallic state is not caused by the introduction of a ``band'' crossing.

\medskip

\paragraph{Charge Carrier Concentration}

\begin{figure}[htp]
\centering
\includegraphics[angle=-90,width=0.45\textwidth]{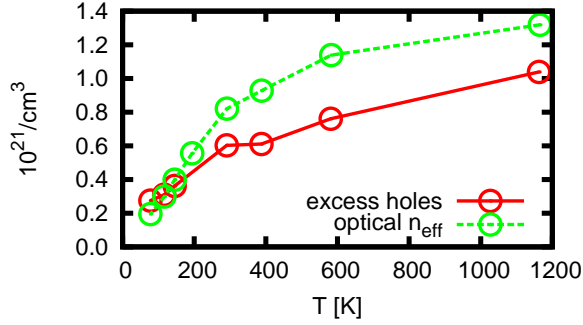}
\caption{{\bf Temperature dependence of the carrier concentration.} Density of carriers extracted from the one-particle spectrum (red) and from the optical conductivity (green). See text for details.}
\label{npart}
\end{figure}

As a reference, we further provide, in \fref{npart}, the temperature dependence of the carrier concentration, extracted from our data by two different means~:\\
\noindent
{\it (1) from the spectral function.} Using $n=\oo{V}\int_0^\infty d\omega f(\omega) A(\omega)$ and
$p=\oo{V}\int_{-\infty}^0 d\omega f(-\omega) A(\omega)$ with the unit-cell volume $V$ and the local spectral function $A(\omega)$ we compute the number of electrons and holes, respectively.
We find that the net carriers are holes at all temperatures ($p-n>0$).
This is to be contrasted with the Seebeck coefficient that from low to high temperatures changes from p-type to n-type and back to p-type, demonstrating that particle-hole symmetry for different response functions can be distinct.

\noindent
{\it (2) from the restricted sum-rule of the optical conductivity.} Using $\int_0^{\Delta}d\omega\sigma(\omega)=\frac{\pi e^2}{2m^*}n_{eff}$ with a gap $\Delta=60meV$ and the orbital
averaged effective mass $m^*=1.5m$, we find values for the number of effective carriers that are in good agreement with those obtained from the one-particle spectrum.

We however note that, in particular at low temperatures, extrinsic effects that are not included in our calculation can potentially have a large effect on the number of carriers.
Indeed the sign of the Hall coefficient seems to be notably sample dependent.

\medskip

\paragraph{Dependence of Results on the Hund's coupling.} 

\begin{figure*}[h!t]
  \begin{center}
\includegraphics[clip,angle=0,width=0.95\textwidth]{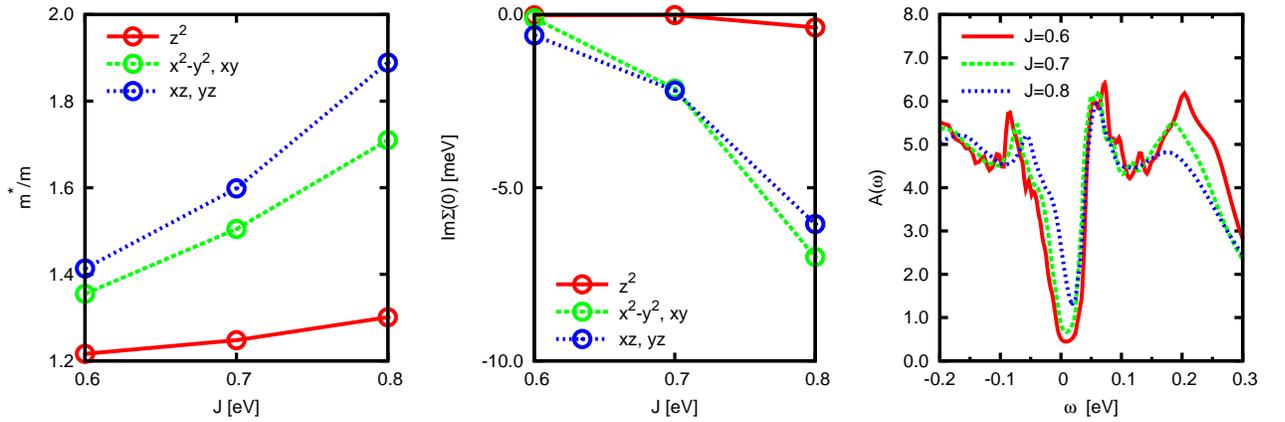}
      \caption{Dependence of the effective masses (left), the imaginary parts of the self-energy (middle) and the spectrum (right) on the Hund's rule $J$, for $U=5.0$eV and $T=116$K.}
      \label{figSUP_J}
      \end{center}
\end{figure*}

In FeSi there is no direct competition between the Hund's coupling $J$ and the crystal field splittings, as e.g.\ manifest from the energy difference between the poles of the impurity Greens 
function (see above). The latter is about 0.3eV, while the Hund's coupling used for our calculations is more than twice as large~: $J=0.7$eV.
Still, there is a notable dependence on the value of $J$ (see also the recent Ref. \cite{0295-5075-95-4-47007})~: In \fref{figSUP_J} (a) are displaced the effective masses $m^*/m$ as obtained from
the slope of the local self-energy at the Fermi level, as well as (b) the imaginary part of the self-energy at the Fermi level.

As was found previously for other iron based materials\cite{1367-2630-11-2-025021,Yin_pnictide} as well as 4d systems\cite{PhysRevLett.106.096401}, the effective masses increase significantly with $J$. The charge gap as shown in \fref{figSUP_J} (c) shrinks accordingly.
The general mechanism and systematics behind this are investigated e.g.\ in Ref.\cite{PTP.49.1483,PhysRevLett.106.096401,Yin_pnictide,PhysRevLett.106.096401,PhysRevB.83.205112,PhysRevLett.107.256401},
and can be explained by resorting to a description in terms of a self-consistent Kondo problem\cite{PTP.49.1483}.
A large Hund's rule coupling constraints electrons to same spin, different orbital states, thereby ``orbitally blocking''\cite{Yin_pnictide} the Kondo interaction,
which in turn decreases the coherence temperature exponentially\cite{PTP.49.1483}.
As a consequence, also the imaginary part of the self-energy, as shown in \fref{figSUP_J} (b), grows significantly in magnitude when increasing $J$. 
For our main calculations, we use the value $J=0.7$eV which is found to be typical for iron based compounds\cite{PhysRevB.82.045105}.

\medskip

\paragraph{Spin state.}

\begin{figure*}[th]
  \begin{center}
\subfigure[\,   T=116K]{\includegraphics[angle=-90,width=.48\textwidth]{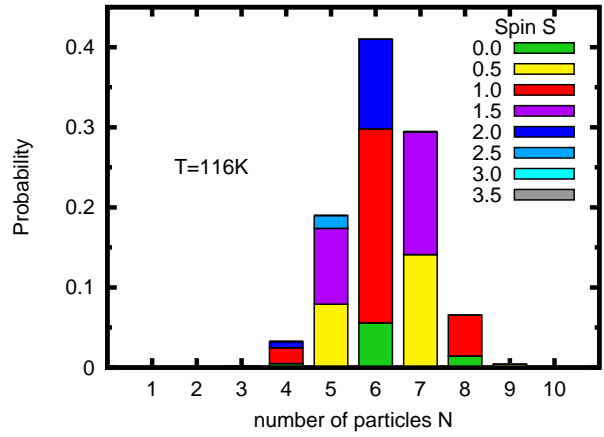}}
\subfigure[\,   T=1116K]{\includegraphics[angle=-90,width=.48\textwidth]{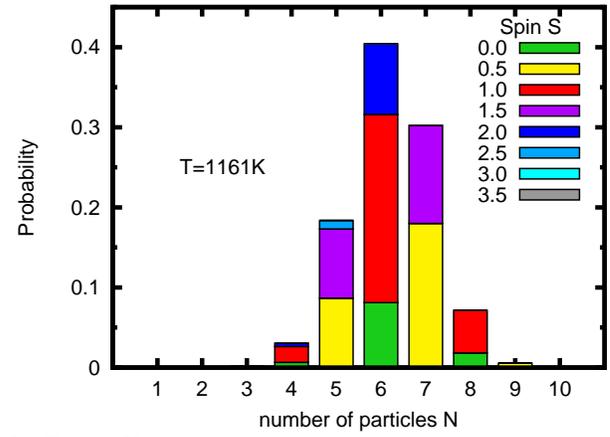}}
      \caption{Probability of atomic states of the DMFT impurity, decomposed in number of particles $N$ (x-axis) and spin-state $S$ (colors). Results for (a) $T=116K$ and (b) $T=1161K$.}
      \label{figSUP1}
      \end{center}
\end{figure*}

The figures \fref{figSUP1}(a) and (b) contain the histograms of the ctqmc\cite{PhysRevB.75.155113} for two different temperatures~: Shown is the probability distribution of the many-body wave-function onto the eigenstates of the effective iron atom representing the 3d orbitals, decomposed into the number of particles $N$ and the spin state $S$.
The main conclusions that we can draw from these results are
\begin{itemize}
	\item the decomposition onto $N$ and $S$ is almost independent of temperature, but strong fluctuations at short time-scales are present.
	\item the system is strongly mixed valence, since the charge variance is of order 1~: $\delta N=\langle\left( N-\langle N\rangle\right)^2\rangle\approx 0.93$.
	\item the spin state S=1 has the overall largest probability at all temperatures, yielding an effective moment
	$M=\sqrt{S(S+1)}g_s\approx3$ ($g_s=2$), with variance $\delta S=\langle\left( S-\langle S\rangle\right)^2\rangle\approx 0.33$.
\end{itemize}
Thus, while the variances of $S$ and $N$ are large, they do not evolve with temperature. This is in strong contrast to systems like MnO, or LaCoO$_3$ in which spin-state transitions occur
\cite{kunes_mno,jmt_mno,PhysRev.155.932,PhysRevLett.106.256401}. 

We note that without averaging over states with the same S, the single most probable state has ($N$,$S$)=(7,1.5), closely followed by ($N$,$S$)=(6,2). This is similar to
other iron systems like e.g. the pnictide BaFe$_2$As$_2$\cite{Yin_pnictide}. However, in BaFe$_2$As$_2$ the most prominent states have a probability that is larger by a factor of 4 than most other states \cite{Yin_pnictide}, while the distribution of states of FeSi is remarkably uniform, as indicated by the character of the state averaged $S$.




\end{document}